\documentclass[aps,twocolumn,prl,superscriptaddress,showpacs,floatfix]{revtex4}
\usepackage{multirow}
\usepackage{graphicx}
\usepackage{xspace}
\usepackage{amssymb}

\newcommand{\axi}{$\overline{\Xi}^+$}
\newcommand{\xim}{$\Xi^-$}
\newcommand{\aom}{$\overline{\Omega}^+$}
\newcommand{\omm}{$\Omega^-$}
\newcommand{\sqrts}{$\sqrt{s_{_{NN}}}$}
\newcommand{\pT}{$\it p_{T}$\xspace}
\newcommand{\Npart}{$\langle N_{part}\rangle$\xspace}
\newcommand{\Nbin}{$\langle N_{bin}\rangle$\xspace}
\newcommand{\pp}{{\it p+p}\xspace}

\begin{document}

\title{Enhanced strange baryon production in Au+Au collisions compared
to \pp at \sqrts = 200 GeV}

\affiliation{Argonne National Laboratory, Argonne, Illinois 60439}
\affiliation{University of Birmingham, Birmingham, United Kingdom}
\affiliation{Brookhaven National Laboratory, Upton, New York 11973}
\affiliation{California Institute of Technology, Pasadena, California 91125}
\affiliation{University of California, Berkeley, California 94720}
\affiliation{University of California, Davis, California 95616}
\affiliation{University of California, Los Angeles, California 90095}
\affiliation{Universidade Estadual de Campinas, Sao Paulo, Brazil}
\affiliation{Carnegie Mellon University, Pittsburgh, Pennsylvania 15213}
\affiliation{University of Illinois at Chicago, Chicago, Illinois 60607}
\affiliation{Creighton University, Omaha, Nebraska 68178}
\affiliation{Nuclear Physics Institute AS CR, 250 68 \v{R}e\v{z}/Prague, Czech Republic}
\affiliation{Laboratory for High Energy (JINR), Dubna, Russia}
\affiliation{Particle Physics Laboratory (JINR), Dubna, Russia}
\affiliation{University of Frankfurt, Frankfurt, Germany}
\affiliation{Institute of Physics, Bhubaneswar 751005, India}
\affiliation{Indian Institute of Technology, Mumbai, India}
\affiliation{Indiana University, Bloomington, Indiana 47408}
\affiliation{Institut de Recherches Subatomiques, Strasbourg, France}
\affiliation{University of Jammu, Jammu 180001, India}
\affiliation{Kent State University, Kent, Ohio 44242}
\affiliation{University of Kentucky, Lexington, Kentucky, 40506-0055}
\affiliation{Institute of Modern Physics, Lanzhou, China}
\affiliation{Lawrence Berkeley National Laboratory, Berkeley, California 94720}
\affiliation{Massachusetts Institute of Technology, Cambridge, MA 02139-4307}
\affiliation{Max-Planck-Institut f\"ur Physik, Munich, Germany}
\affiliation{Michigan State University, East Lansing, Michigan 48824}
\affiliation{Moscow Engineering Physics Institute, Moscow Russia}
\affiliation{City College of New York, New York City, New York 10031}
\affiliation{NIKHEF and Utrecht University, Amsterdam, The Netherlands}
\affiliation{Ohio State University, Columbus, Ohio 43210}
\affiliation{Panjab University, Chandigarh 160014, India}
\affiliation{Pennsylvania State University, University Park, Pennsylvania 16802}
\affiliation{Institute of High Energy Physics, Protvino, Russia}
\affiliation{Purdue University, West Lafayette, Indiana 47907}
\affiliation{Pusan National University, Pusan, Republic of Korea}
\affiliation{University of Rajasthan, Jaipur 302004, India}
\affiliation{Rice University, Houston, Texas 77251}
\affiliation{Universidade de Sao Paulo, Sao Paulo, Brazil}
\affiliation{University of Science \& Technology of China, Hefei 230026, China}
\affiliation{Shanghai Institute of Applied Physics, Shanghai 201800, China}
\affiliation{SUBATECH, Nantes, France}
\affiliation{Texas A\&M University, College Station, Texas 77843}
\affiliation{University of Texas, Austin, Texas 78712}
\affiliation{Tsinghua University, Beijing 100084, China}
\affiliation{Valparaiso University, Valparaiso, Indiana 46383}
\affiliation{Variable Energy Cyclotron Centre, Kolkata 700064, India}
\affiliation{Warsaw University of Technology, Warsaw, Poland}
\affiliation{University of Washington, Seattle, Washington 98195}
\affiliation{Wayne State University, Detroit, Michigan 48201}
\affiliation{Institute of Particle Physics, CCNU (HZNU), Wuhan 430079, China}
\affiliation{Yale University, New Haven, Connecticut 06520}
\affiliation{University of Zagreb, Zagreb, HR-10002, Croatia}

\author{B.I.~Abelev}\affiliation{University of Illinois at Chicago, Chicago, Illinois 60607}
\author{M.M.~Aggarwal}\affiliation{Panjab University, Chandigarh 160014, India}
\author{Z.~Ahammed}\affiliation{Variable Energy Cyclotron Centre, Kolkata 700064, India}
\author{B.D.~Anderson}\affiliation{Kent State University, Kent, Ohio 44242}
\author{D.~Arkhipkin}\affiliation{Particle Physics Laboratory (JINR), Dubna, Russia}
\author{G.S.~Averichev}\affiliation{Laboratory for High Energy (JINR), Dubna, Russia}
\author{Y.~Bai}\affiliation{NIKHEF and Utrecht University, Amsterdam, The Netherlands}
\author{J.~Balewski}\affiliation{Indiana University, Bloomington, Indiana 47408}
\author{O.~Barannikova}\affiliation{University of Illinois at Chicago, Chicago, Illinois 60607}
\author{L.S.~Barnby}\affiliation{University of Birmingham, Birmingham, United Kingdom}
% ...

\collaboration{STAR Collaboration}\noaffiliation

\begin{abstract}

We report on the observed differences in production rates of strange
and multi-strange baryons in Au+Au collisions at \sqrts = 200 GeV
compared to \pp interactions at the same energy. The strange baryon yields in Au+Au collisions,
 when scaled down by the number of participating nucleons, are enhanced relative to those
measured in \pp reactions.  The enhancement  observed
increases with the strangeness content of the baryon, and increases for
all strange baryons with collision centrality.   The enhancement is
qualitatively similar to that observed at lower collision energy
\sqrts =17.3 GeV. The previous observations are for the bulk
production, while at intermediate \pT, 1 $<{\it p_{T}}<$ 4 GeV/c, the strange baryons even
exceed binary scaling from \pp yields.

\end{abstract}

 \maketitle

One of the aims of studying relativistic heavy ion collisions is to observe how
matter behaves at extremes of temperature and/or density. The energy
densities in the medium  produced by these collisions are far from
that of ground state nuclear matter. Ultimately we hope to determine
if they are sufficiently high to create a system where the degrees
of freedoms are those of quarks and gluons, a state called the
Quark-Gluon Plasma (QGP). By comparing the particles produced in A+A
to those from \pp collisions, in which a QGP phase is not expected,
we can gain insight into the properties of the medium.

 Strange particles are of particular interest
 since the initial strangeness content of the colliding nuclei is very small
 and there is no net strangeness. This means that all strange hadrons must be formed in the  matter
  produced. Originally, it was proposed that strangeness production would be increased due to
  the formation of  a QGP compared to that from a hadron gas~\cite{RafEnhance}. This
enhancement is due to the high production rate of {\it gg}
$\rightarrow$ {\it s}$\bar{s}$ in a QGP, a process absent in the
hadronic state. The subsequent hadronization of these (anti)strange
quarks results in a significant increase in strange particle
production, thus signaling a plasma was formed.

The concept of enhanced strangeness production in the QGP can be
recast in the language  of statistical mechanics. A Grand Canonical
Ensemble limit is likely only to be reached in the high multiplicity
of heavy ion reactions. If this is the case, any measured
enhancement is really a phase space suppression in \pp reactions
that is removed in the heavy ion case. This lack of available phase
space in small systems, such as those from \pp collisions, requires
a Canonical ensemble to be used which results in  a suppression of
strangeness production when scaled to the appropriate
volume~\cite{Redlich,Becattini}. However, there is no a priori
method for directly calculating this volume
  and thus the authors make the simplest hypothesis and assume that the volume is linearly proportional to
  the number of collision participants, \Npart. The degree of suppression  increases with the
strange
 quark content of the particle. For sufficiently
 large volumes, the system is thermalized, the phase space suppression effects
 disappear,  and the yields scale linearly with the volume, i.e. \Npart.
 Initial measurements from the SPS suggested such a linear \Npart scaling~\cite{NA57Enhance}.
  However, it is not observed at RHIC~\cite{ScalingPaper} or in the more recent SPS
  results~\cite{NA57New}.

 While the observables mentioned above are sensitive to the bulk of
the produced particles with
 momenta below 2 GeV/c, further important information can be extracted from
intermediate and high \pT particles. At RHIC, hadrons are suppressed
at intermediate to high \pT when compared to binary-scaled \pp data
at the same energy~\cite{RaaNch200}. This effect is attributed to
the energy loss of partons as they traverse the hot and dense medium
produced~\cite{JetQuenchTheory1,JetQuenchTheory2}. Measurements
using identified particles help shed light on the details of the
energy loss mechanism.

In this paper we present further analysis of the high statistics
measurements from {\it p+p} and Au+Au collisions at \sqrts=200 GeV
for strange and multi-strange
 baryon production at mid-rapidity as reported by the STAR collaboration at
RHIC~\cite{ppPaper,ScalingPaper}. Details of the STAR experiment are
in \cite{STAR}. Specific details of the trigger and detectors used
to collect the data reported here can be found in
~\cite{ppPaper,ScalingPaper} and references therein. The Au+Au event
sample consisted of 1.5$\times$10$^{6}$ central collision triggers
and 1.6$\times$10$^{6}$ minimum bias triggers. The \pp results are
from 6$\times$10$^{6}$ minimum bias events. Particle identification
is via the reconstruction of the charged daughter decay particles in
the Time Projection Chamber. The decay channels used are  $\Lambda
\rightarrow {\it p }+ \pi^{-}$, $\Xi^{-} \rightarrow \Lambda +
\pi^{-} \rightarrow {\it p} + \pi^{-} + \pi^{-} $ and $\Omega^{-}
\rightarrow \Lambda + K^{-} \rightarrow {\it p} + \pi^{-} + K^{-}$
plus the charge conjugates for the anti-particle decays.

After  cuts, to reduce random combinatorics, parent particles were
selected if the calculated invariant mass fell within 3$\sigma$
around the peak after background subtraction. The data were
corrected, as a function of \pT, for efficiency and detector
acceptance. Monte-Carlo studies showed that the
 corrections were constant as a function of rapidity
 over the measured regions. Further details of these reconstruction
 and correction techniques can be found in \cite{ScalingPaper,ppPaper} and references therein.
 Several contributions to the systematic uncertainty of particle
yields were studied: detector simulation and efficiency
calculations, inhomogeneities of the detector responses, pile-up
effects and the extrapolation of the data fits to zero \pT. In \pp
collisions an additional normalization error due to varying beam
luminosity and trigger efficiencies of $\sim 4 \%$ is included. The
$\Lambda$ yields were corrected for feed-down from multi-strange
baryons using the measured spectra, the correction was of the order
of 15$\%$.

\begin{figure}[htb]
\begin{center}
\includegraphics[width=0.45\textwidth]{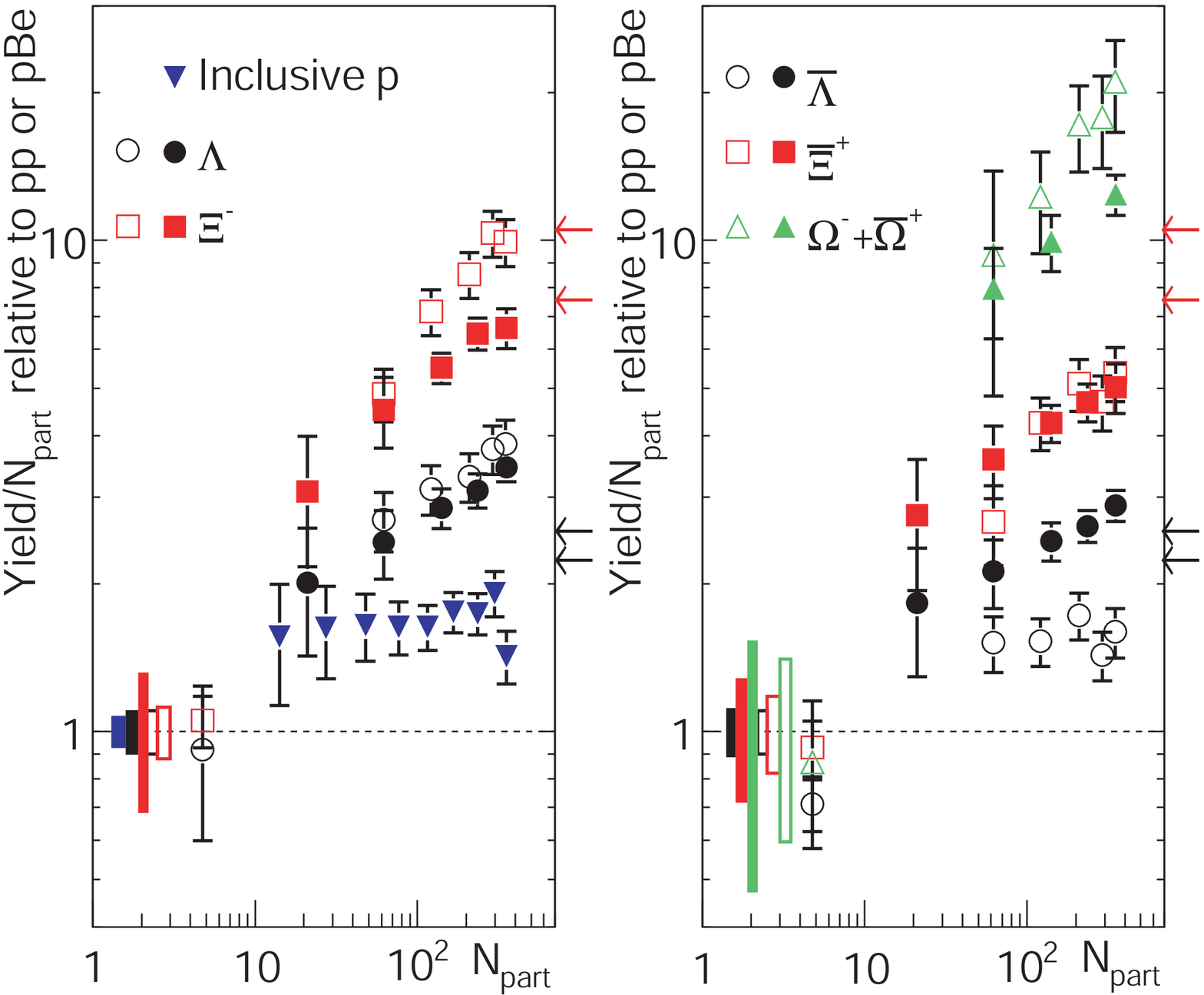}
\caption{ (color online) Mid-rapidity {\it E(i)} as a function of \Npart for
$\Lambda$, $\bar{\Lambda}$ ($|$y$|$ $<$ 1.0), \xim, \axi, \omm +
\aom ($|$y$|$ $<$ 0.75) and inclusive p ($|$y$|$ $<$ 0.5). Boxes at
unity show statistical and systematical uncertainties combined in
the \pp (p+Be) data. Error bars on the data points represent those
from the heavy-ions. The solid markers are for Au+Au at \sqrts=200
GeV and the open symbols for Pb+Pb ($|$y$|$ $<$ 0.5) at \sqrts=17.3
GeV~\cite{NA57Enhance}. The arrows on the right axes mark the
predictions from a GC formalism model when varying {\it T} from 165 MeV
({\it E(\xim)}=10.7, {\it E($\Lambda$)}=2.6) to 170 MeV ({\it E(\xim)}=7.5,
{\it E($\Lambda$)}=2.2). The red arrows indicate the predictions for $\Xi$ and
the black arrows those for $\Lambda$ , see text for
details~\cite{RedlichPrivate}. } \label{Fig:Enhancement}
\end{center}
\end{figure}

For each species, {\it i}, the yield enhancement, {\it E(i)}, above
that expected from \Npart scaling was calculated using:
\begin{equation}
 E(i) =\frac{ Yield^{AA}(i)\langle N_{part}^{NN}\rangle}{
Yield^{NN}(i)\langle N_{part}^{AA}\rangle}
\end{equation}
Fig.~\ref{Fig:Enhancement}  shows {\it E(i)} as a function of
\Npart, the inclusive proton data illustrate the effects for
non-strange baryons~\cite{pbarPaper}. Mid-rapidity hyperon yields
measured as a function of centrality in Au+Au~\cite{ScalingPaper}
and \pp~\cite{ppPaper} collisions were used. The number of
participants, \Npart, and the number of binary collisions, \Nbin,
were estimated via a Monte-Carlo Glauber
calculation~\cite{Glauber,StarGlaub}. Since the \pp data were
recorded with a trigger that was only sensitive to the non-singly
diffractive (NSD) part of the total inelastic cross-section, all \pp
yields have been corrected by
$\sigma^{NN}_{NSD}$/$\sigma^{NN}_{inel}$, (=30/42), to obtain the
total invariant cross-sections.

It can be seen that there is an enhancement in the yields over that
expected from \Npart scaling for all the particles presented. Since
the proton yields are not corrected for feed-down, which is
predominantly from the $\Lambda$ and $\Sigma$, the  {\it p}
measurement is actually a sum of the  primary protons and those from
secondary decays. The integrated $\Lambda+\Sigma^{0}$ over inclusive
{\it p} ratio varies from 30 to 40 $\%$ for the \pp and Au+Au
collisions respectively. If only primary protons were measured {\it
E(proton)} would be closer to unity. A hierarchy in the scale of
enhancements, which grows with increased strangeness of the baryon,
is observed. This trend is predicted by Grand Canonical (GC)
ensemble approaches, as is the fact that the E(i) values for each
baryon/anti-baryon pair are similar in shape~\cite{Redlich}. The
difference in the scale of the enhancements for baryon and
anti-baryon, especially at the SPS, is due to the existence of a
non-zero net-baryon number. However, the ratio of {\it
E(anti-baryon)} to {\it E(baryon)} varies as a function of \Npart at
the SPS, possibly signifying  different production/annhilation
mechanisms for (anti)particles at the SPS compared to at RHIC. For
instance the net-$\Lambda$ yields at the SPS, can be successfully
described via multiple interactions of the projectile
nuclei~\cite{AGSEnhance}. This effect is expected to be less
significant at RHIC. It is also interesting to note that the
measured enhancements for the $\Lambda$, anti($\Xi$),
 and $\Omega$ at RHIC are the same, within errors, as those calculated from the mid-rapidity SPS
data (open symbols in Fig.~\ref{Fig:Enhancement}) despite an order
of magnitude increase in the collision energies. Theoretical
predictions using the GC ensemble approach predict a significant
decrease in all the (anti)baryon enhancements with collision
energy~\cite{Redlich}. A GC model, with a chemical freeze-out
temperature of {\it T}=165 MeV and a baryon chemical potential,
$\mu_{b}$ =29 MeV calculates enhancements of {\it E(\xim)}=10.7 and
{\it E($\Lambda$)}=2.6 for the most central Au+Au events at
\sqrts=200 GeV~\cite{RedlichPrivate}. These enhancement calculations
cannot consistently describe the (anti)$\Xi$ and the (anti)$\Lambda$
enhancements.
  However, the
scales of the enhancements are very sensitive to the assumed
freeze-out temperature and if {\it T}=170 MeV is used {\it
E(\xim)}=7.5 and {\it E($\Lambda$)}=2.2.

 While the measured enhancements are approximately constant for
 the inclusive protons they are clearly not for the $\Lambda$, $\Xi$, and $\Omega$;
 this is again counter to theoretical expectations where
 the dependence of the strange baryon yields is expected to be
 linear with \Npart for \Npart $\gtrsim$ 20. One explanation
for this deviation from the theory  is that the volume responsible
for  strangeness production is not linearly proportional to the
geometrical overlap region, as assumed in the model. A model that
gives a reasonable description of the magnitudes and shapes of the
enhancements with respect to centrality is described in
\cite{RafOverSat}. This model allows for an over-saturation of
strange quarks, which varies with centrality, and thus does not
invoke chemical equilibration.

\begin{figure}[htb]
\begin{center}
\includegraphics[width=0.45\textwidth]{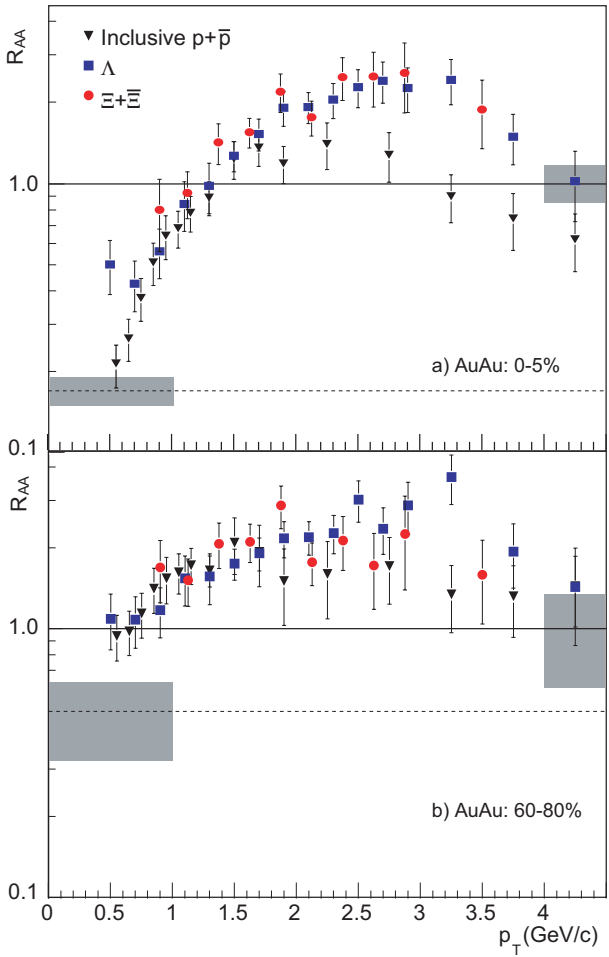}
\caption{(Color Online) $R_{AA}$ from a) 0-5$\%$ and b) 60-80\% central Au+Au
events for p+$\bar{{\rm p}}$~\cite{TofProtonspp,TofAA}, $\Lambda$
and \xim+\axi. Errors shown are statistical plus systematic added in
quadrature. The band at unity shows the systematical uncertainty on
\Nbin. The dashed line below unity shows the expected value of
R$_{AA}$ should the yields scale with \Npart and the band around it
shows the systematic uncertainty on \Npart.} \label{Fig:Raa}
\end{center}
\end{figure}

Fig.~\ref{Fig:Enhancement} is an average measurement of the
difference in production between nucleus-nucleus and nucleon-nucleon
collisions. Since the \pT distributions of the particles are
approximately exponential these results are dominated by the physics
occurring at \pT$ \lesssim$ 2 GeV/c. Differences in the \pT
distributions for \pp and Au+Au data are studied by calculating the
nuclear modification factor:
\begin{equation}
 R_{AA}(p_{T},i)
 =\frac{d^{2}N^{AA}(i)/dp_{T}dy}{T_{AA}d^{2}\sigma^{NN}(i)/dp_{T}dy},
\end{equation}
 $T_{AA}$ = \Nbin/$\sigma^{NN}_{inel}$. Fig.~\ref{Fig:Raa}a shows R$_{AA}$ for $\Lambda$ and
 the sum $\Xi$+$\bar{\Xi}$ for
0-5$\%$ Au+Au collisions along with those for inclusive p+$\bar{{\rm
p}}$ measurements~\cite{TofProtonspp,TofAA}.

\begin{figure}[tb]
\begin{center}
\includegraphics[width=0.45\textwidth]{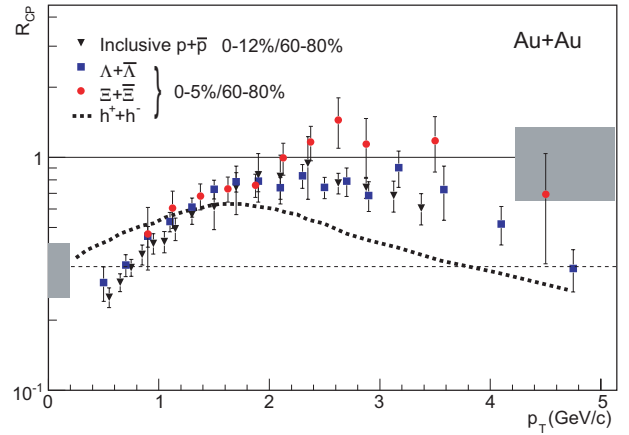}
\caption{(Color online) $R_{CP}$ Au+Au events for p+$\bar{{\rm p}}$
0-12$\%$/60-80$\%$~\cite{TofAA}, $\Lambda$+$\bar{\Lambda}$ and \xim+\axi
0-5$\%$/60-80$\%$~\cite{ScalingPaper}. Also shown as the dashed
curve are the results for h$^{+}$+h$^{-}$ for 0-5$\%$/60-80$\%$~\cite{Rcph}.
Errors shown are statistical plus systematic added in quadrature.
The band at unity shows the systematical uncertainty on \Nbin. The
dashed line below unity shows the expected value of R$_{CP}$ should
the yields scale with \Npart and the band around it shows the
systematic uncertainty on \Npart.} \label{Fig:Rcp}
\end{center}
\end{figure}

A striking feature of Fig.~\ref{Fig:Raa} is that both the central
(top panel) and peripheral (bottom panel) R$_{AA}$ distributions for
the $\Lambda$ and \xim+\axi reach maxima that are much greater than
unity, a value that would signify binary collision scaling. In fact
the peripheral collision R$_{AA}$ distributions for the hyperons,
Fig.~\ref{Fig:Raa}b, are of approximately the same magnitude as the
central R$_{AA}$ data, Fig.~\ref{Fig:Raa}a, at intermediate to high
\pT. These results are in contrast to the earlier reported
suppression of high \pT hyperons observed via
$R_{CP}$~\cite{TofAA,RcpPID,ScalingPaper,Rcph}, these data are
reproduced in Fig.~\ref{Fig:Rcp}.
\begin{equation}
 R_{CP}(p_{T},i)
 =\frac{[d^{2}N^{cent}(i)/dp_{T}dy/\langle N_{bin}^{cent} \rangle]}{[d^{2}N^{periph}(i)/dp_{T}dy/\langle N_{bin}^{periph}\rangle
]},
\end{equation}
Non-strange hadrons reveal a similar suppression  when using  \pp or
peripheral Au+Au collisions as a reference. For \pT $ > $ 1.5 GeV/c,
unidentified charged hadrons show a suppression of the Au+Au
spectra~\cite{RaaNch200}. Comparing R$_{AA}$, Fig.~\ref{Fig:Raa}, to
R$_{CP}$, Fig.~\ref{Fig:Rcp}, shows that R$_{AA}$($\Lambda$)
$\approx$ R$_{AA}$($\Xi$) $\neq$ R$_{AA}$(p) but that
R$_{CP}$($\Lambda$) $\approx$ R$_{CP}$($\Xi$) $\approx$ R$_{CP}$(p),
especially at intermediate to high \pT. This is possibly due to
phase space effects in the \pp data extending to this intermediate
\pT regime. It is surprising that this decreased production in \pp
events, while predicted in the soft physics/thermal production
regime, i.e. \pT $<$ 2 GeV/c, extends out to, and even dominates in,
this intermediate \pT region. Fig.~\ref{Fig:Raa}a suggests that this
effect is strong out to \pT$\sim$ 3 GeV/c. The shapes of the
$R_{CP}$ distributions at intermediate to high \pT are generally
interpreted as the result of parton energy loss in the hot dense
matter and quark coalescence during hadronization. A comparison of
Fig.~\ref{Fig:Raa}a and Fig.~\ref{Fig:Raa}b shows that the turnover
points  occur at approximately the same \pT. These data suggest that
an enhancement of strangeness production has already set in in
peripheral Au+Au collisions. This behavior is similar to that
observed for the total yields in Fig.~\ref{Fig:Enhancement}, and
quantitatively consistent with expectations from canonical
suppression in \pp.  Some portion of the R$_{AA}$ peak may be
explained via the Cronin effect, the observed increase in
intermediate \pT spectra in {\it p}-A collisions~\cite{Cronin}.
However, the Cronin enhancement stays constant, or possibly
increases, as a function of centrality~\cite{TofProtons}, and this
is not seen in our data. Effects due to radial flow in the Au+Au
data are significant at RHIC energies, even for the multi-strange
baryons~\cite{StarFlowXi}, but flow dominates only at low \pT. The
shapes of the R$_{AA}$ distributions below 1 GeV/c are markedly
different. The peripheral collision data indicate approximate binary
scaling of the baryon yields while the most central data fall
beneath binary scaling but significantly above that suggesting
participant scaling. This again indicates that there are different
constraints on baryon production when going from \pp to peripheral
to central Au+Au collisions.

\begin{figure}[tb]
\begin{center}
\includegraphics[width=0.45\textwidth]{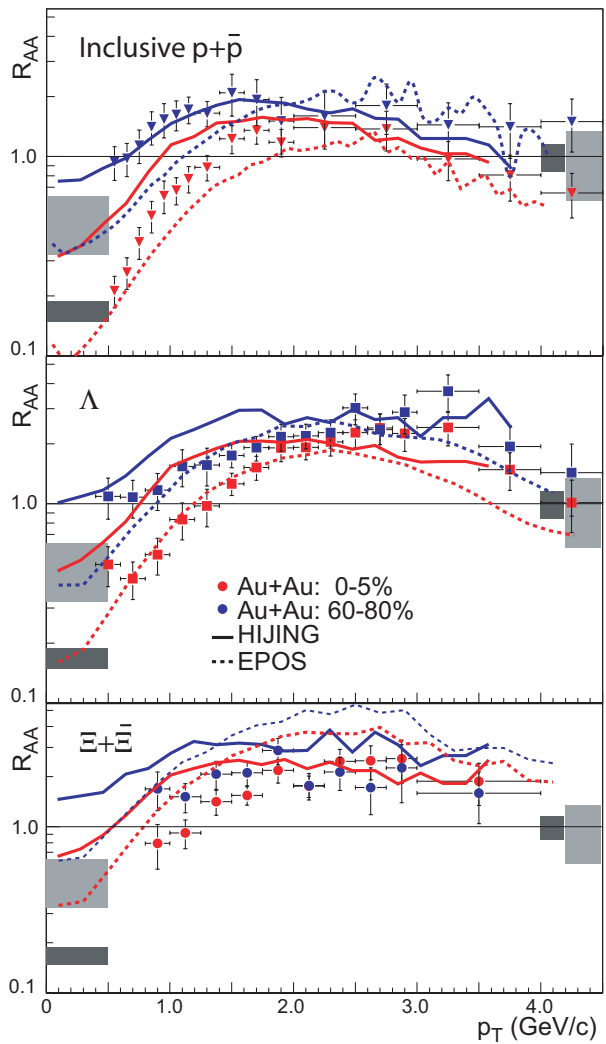}
\caption{(Color online) Comparison of $R_{AA}$ from data to HIJING~\cite{HIJINGcalc} and EPOS~\cite{EPOScalc} for
0-5$\%$ and 60-80$\%$ most central Au+Au collisions, for  p+$\bar{{\rm p}}$
, $\Lambda$ and \xim+\axi. Errors shown are statistical plus systematic added in quadrature.
The bands at unity shows the systematical uncertainty on \Nbin. The
bands below unity on the left of the graphs are centered at the expected value of R$_{AA}$ should
the yields scale with \Npart and the widths of the bands indicate the
systematic uncertainty on \Npart.} \label{Fig:RAATheory}
\end{center}
\end{figure}

Comparisons to dynamical models can be used to understand  in more
detail how the close-to-equilibrium strangeness production can be
achieved and whether the same mechanisms affect strange particle
production at intermediate \pT. In the HIJING model~\cite{HIJING} the yields and
qualitative features of the strange baryon R$_{AA}$ measurements (solid curves in Fig.~\ref{Fig:RAATheory})  can
only be obtained  when baryon junctions and color strings are
included~\cite{HIJINGStrings,HIJINGcalc}. EPOS calculations~\cite{EPOS,EPOScalc} (dashed curves
in Fig.~\ref{Fig:RAATheory}) produce similarly large differences  in the hyperon R$_{AA}$ and
  R$_{CP}$~\cite{EPOS} to those measured at RHIC and also give a qualitatively reasonable
  representation of the shape of the data.  EPOS
 describes particle production via a parton model where Au+Au collisions are represented as many binary
 interactions. Each binary interaction is described by a longitudinal color field which is expressed as
 a relativistic string, or parton ladder. At a very early proper time, before hadronization, the collision region
 is split into two environments, the core, in which the density of strings is high,
and the corona, which surrounds the core and has a low string density. Production from the corona
 is due to collisions of nucleons at the periphery of the nuclei and modeled via string fragmentation. Corona
  production is thus similar to that from \pp collisions. Meanwhile particle production
 from the core is approximated via a simple statistical hadronization process, similar to that described
 in~\cite{StatMod}, a collective flow profile is then imposed upon these particles. The relative weight
 of core to corona production varies with both centrality and
  particle species with the core dominating production in the most central events.
   Strange baryons are dominated by core production even in peripheral events.

In summary, we observe enhanced strange baryon mid-rapidity
production in Au+Au collisions, especially in the more  central
events, when compared to the \Npart scaled \pp data from the same
energy. The measured yields fail to scale with \Npart as predicted
if the GC regime is reached and the particle production volume
scales with the geometrical overlap region. The magnitudes of the
suppressions are different to those predicted, but close to those
measured at SPS energies. At intermediate \pT the R$_{AA}$ values
are higher than binary scaling of \pp data would predict.  When
attempting to understand the evolution of strange particle
production from \pp to central Au+Au one must take into account both
the  effects due to a suppression of strangeness production in
 \pp and jet quenching plus quark recombination in Au+Au collisions, with the former
 dominating at intermediate \pT. Since the measured R$_{CP}$ values for all
strange baryons equal those of the inclusive protons in the
intermediate and high \pT  regions, and are significantly below
binary scaling, the \pp-like suppression is already predominantly
removed in peripheral Au+Au collisions.

We thank the RHIC Operations Group and RCF at BNL, and the
NERSC Center at LBNL and the resources provided by the
Open Science Grid consortium for their support. This work was supported
in part by the Offices of NP and HEP within the U.S. DOE Office
of Science; the U.S. NSF; the BMBF of Germany; CNRS/IN2P3, RA, RPL, and
EMN of France; EPSRC of the United Kingdom; FAPESP of Brazil;
the Russian Ministry of Sci. and Tech.; the Ministry of
Education and the NNSFC of China; IRP and GA of the Czech Republic,
FOM of the Netherlands, DAE, DST, and CSIR of the Government
of India; Swiss NSF; the Polish State Committee for Scientific
Research; Slovak Research and Development Agency, and the
Korea Sci. and Eng. Foundation.

\end{document}